\documentclass[preprint,12pt]{elsarticle}

\usepackage{amssymb}
\usepackage{graphicx}
\usepackage{tempora}                

\usepackage{csquotes}

\usepackage[version=3]{mhchem} 
\usepackage{color}
\usepackage[normalem]{ulem}
\usepackage{dcolumn}
\usepackage{bm}
\usepackage[unicode=true,colorlinks=true,citecolor=blue,urlcolor=blue]{hyperref}

\usepackage{braket}

\journal{Journal of Non-Crystalline Solids}

\begin{document}

\begin{frontmatter}



\title{Spontaneous noise of birefringence in rare-earth doped glasses}


\author[solab]{Kozlov V.~O.}
\author[solab,phot]{Ryzhov I.~I.}
\author[solab,ssd]{Kozlov G.~G.}
\author[itmo]{Kolobkova E.~V.}
\author[solab]{Zapasskii V.~S.}

\affiliation[solab]{organization={Spin Optics Laboratory, Faculty of Physics, St Petersburg State University},
            addressline={Peterhof, Ul'yanovskaya ul., 1}, 
            city={Saint Petersburg},
            postcode={198504}, 
            country={Russia}}
\affiliation[phot]{organization={Photonics Department, Faculty of Physics, St Petersburg State University},
            addressline={Peterhof, Ul'yanovskaya ul., 1}, 
            city={Saint Petersburg},
            postcode={198504}, 
            country={Russia}}
\affiliation[ssd]{organization={Solid State Department, Faculty of Physics, St Petersburg State University},
            addressline={Peterhof, Ul'yanovskaya ul., 1}, 
            city={Saint Petersburg},
            postcode={198504}, 
            country={Russia}}
\affiliation[itmo]{organization={ITMO University},
            addressline={49 Kronverskii Av.}, 
            city={Saint Petersburg},
            postcode={197101}, 
            country={Russia}}

\begin{abstract}
We report on first direct observation of spontaneous fluctuations of birefringence in glasses doped with rare-earth (RE) ions. The fluctuations were observed in Nd$^{3+}$- and Yb$^{3+}$-doped glasses as polarization noise of the laser beam transmitted through the sample in the region of the RE-ion absorption. The noise was characterized by a flat (``white'') spectrum in the range of frequencies up to 1 GHz and did not show any dependence on magnetic field. The discovered polarization noise is interpreted in terms of structural dynamics of glasses revealed at low temperatures and usually described in the model of tunneling two-level systems (TLS). High sensitivity of the polarization noise technique to this dynamics is related to small homogeneous width of \textit{f-f} transitions of RE-ions in glasses and small spectral width of the probe laser light. The discovered effect provides a new experimental approach to studying low-temperature structural dynamics of different disordered matrices and interactions of impurities with environment in such media.
\end{abstract}

\begin{graphicalabstract}
\centering
\includegraphics[width=\textwidth]{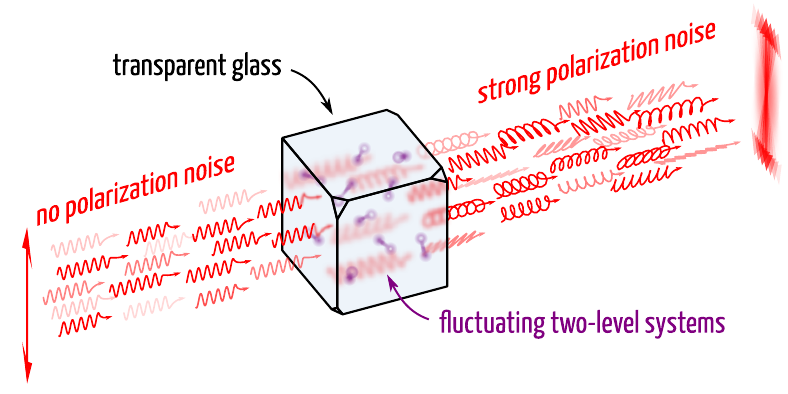}
\end{graphicalabstract}



\end{frontmatter}

\section{Introduction}
The present-day noise spectroscopy as a tool of research includes, along with traditional studies of {obstructive noises which impend the measurements, a more constructive} investigations of noises capable of providing meaningful information about the system under study \cite{vanderziel,physicsofnoise}. One of the methods of this kind that has rapidly developed for the last two decades is the so-called Spin Noise Spectroscopy (SNS) aimed at polarimetric detection of spontaneous spin precession \cite{zapreview,sinitsyn,smirnov}. This experimental technique, first demonstrated on an atomic system \cite{al-zap}, is widely used nowadays for studying magnetic resonance and spin dynamics in semiconductor structures \cite{muller,hubner,hubner1,poltavtsev-sns-single-qw14,ryzhov-sn-explores-local-fields16} and has been recently applied to dielectric crystals with paramagnetic impurities \cite{kamenskii,kamenskii1} {and to birefringent ones in particular~\cite{kozlov-sn-birefringent22}}. The method implies detection of magnetization noise of a paramagnet by measuring fluctuations of its Faraday rotation and allows one to get information about spin system traditionally obtained using the EPR spectroscopy. It is important that measurements of the Faraday rotation noise are usually performed in the region of transparency, where the probe light does not excite the system. This is why the SNS was often considered as a nonperturbative method of EPR spectroscopy. In recent years, however, it was shown that nonperturbativity of the SNS is far from being its main merit and that, under conditions of resonant probing, the SNS may provide valuable information inaccessible for nonresonant measurements. In particular, application of SNS to crystals doped with rare-earth (RE) ions \cite{kamenskii} proved to be possible only due to specific features of this method revealed under {\it resonant probing} of inhomogeneously broadened optical transitions. 

Successful application of SNS to crystals with RE impurities has led us to the idea of its application to RE-doped glasses, with huge predominance of inhomogeneous width of the intraconfigurational (4\textit{f}-4\textit{f}) optical transitions over homogeneous at liquid-helium temperatures \cite{shelby,macfarlane,vanderzaag,narrowline}. We did not count on obtaining structured magnetic resonance spectra (in the studied frequency range below 1 GHz) because of strong inhomogeneous broadening of the precession peaks. Still, we considered even smooth spectrum of the Faraday-rotation noise to be valuable as a fundamentally new source of information. What we have found, however, essentially differed from what we expected. 
The spectrum of the observed noise was virtually ‘white’ (within the frequency range up to $\sim$1~GHz), practically did not vary with magnetic field, and could be observed only at the wavelengths of certain \textit{f-f} transitions. At the same time, the observed noise, for a number of reasons, could not be ascribed to fluctuations of the spin-related gyrotropy of the glass. 

In this paper, we present results of experimental study of the discovered polarization noise and come to conclusion that this noise is related to intrinsic dynamics of the disordered vitreous matrix usually described in terms of two-level systems \cite{anderson,phil1}. This dynamics gives rise to fluctuations of local environment of the RE impurity and to fluctuations of its optical polarizability tensor. The high sensitivity of the probe beam polarization to these fluctuations, in the vicinity of the \textit{f-f} transitions, is related to their small homogeneous widths (in combination with small spectral width of the probe laser beam).

The paper is organized as follows. 
In Section~I we consider the issue important for this work regarding the relation between the type of fluctuating anisotropy of the medium and the type of polarization noise it generates in the probe light. In Section~II we present experimental part of the work including characterization of the samples, details of the experimental technique, and results of the measurements. Section~III describes specific properties of the polarization noise signal. Interpretation of the discovered noise is presented in Section~IV. Section~V briefly summarize results of the work. 

\section{Light polarization noise versus noise of \textbf{anisotropy}}

The fluctuations of the polarization plane azimuth detected in the SNS, are assumed to be related to the {\it gyrotropy noise} of the medium associated with fluctuations of its magnetization. One can easily see, however, that the same polarization noise may also arise in the absence of any gyrotropy noise when the medium with essentially complex optical polarizability exhibits fluctuations of {\it linear}, rather than circular, anisotropy\footnote{We use here the term {\it optical anisotropy} rather than {\it optical birefringence} to emphasize importance of its imaginary part (usually referred to as {\it dichroism}). The term {\it birefringence} often implies {only} real part of the anisotropy.}. How can be distinguished these two types of fluctuating anisotropy? This question arose, in the present study of rare-earth (RE) doped glasses, when we encountered the polarization noise with characteristics unusual for the SNS. Let us consider this issue in more detail. 

First of all, note that fluctuations of {anisotropy} of the medium do not affect the light polarization when it coincides with one of the normal modes of the fluctuating {anisotropy}. For the fluctuating linear or circular {anisotropy}, they are, respectively, linearly or circularly polarized waves. Thus, by choosing polarization of the incident light coincident with one of the normal modes, we can extinguish the relevant polarization fluctuations of the probe light. Here, we can notice a great difference between the media with circular and linear {anisotropy} that allows one to distinguish their fluctuations. Specifically, fluctuations of gyrotropy (circular {anisotropy}) of the medium should not be observed in the noise of circularly polarized light, while fluctuations of linear {anisotropy} with randomly aligned anisotropy axes do not have a common normal mode and, therefore, cannot be extinguished in any linearly polarized light. This distinction can be clearly 
{seen from representation of light polarization on} the Poincar\'e sphere, where the circular and linear polarizations are depicted by points (poles of the sphere) and by the line (equator of the sphere with a multitude of polarizations), respectively. 

Thus, our qualitative consideration allows us to make the conclusion (which may seem trivial) that the noise of gyrotropy should vanish in a circularly polarized probe light. One more conclusion that may be made from the above consideration is that the conventional spin noise, revealed as the noise of the Faraday rotation, cannot be detected in a circularly polarized probe light. 

The above simple reasoning helped us to identify the nature of the anisotropy fluctuations giving rise to polarization noise in the RE-doped glasses. 

\section{Experimental}

\begin{figure}
 \centering
 \includegraphics[width=0.7\linewidth]{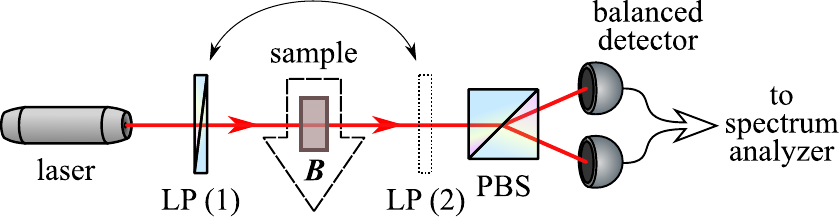}
 \caption{Schematic of the experimental setup. Polarization noise was measured by replacing linear polarizer (LP) from position (1) to position (2). The half-wave plate ($\lambda/2$) was used to balance the polarimetric detector. 
 PBS - polarization beamsplitter. }
 \label{setup} 
 \end{figure}

The measurements were performed on samples of phosphate glasses doped with neodymium (0.6$\cdot$10$^{21}$~cm$^{-3}$) and ytterbium (2$\cdot$10$^{21}$~cm$^{-3}$). The thicknesses of the plates (3 mm and 0.4 mm, respectively) were chosen to provide optical density at resonance below unity. The experimental setup was similar to that commonly used for spin-noise detection (Fig.~\ref{setup}). The sample was placed into a Montana Cryostation cryostat that provided a temperature down to 3~K and magnetic field up to 0.7~T. As a light source we used a tunable CW ring Ti:sapphire laser similar to that employed previously for spin-noise measurements in rare-earth doped crystals~\cite{kamenskii,kamenskii1}. A small width of the laser emission spectrum allowed us to realize the `giant spin-noise gain effect'\cite{kamenskii} on inhomogeneously broadened \textit{f-f} transitions of the RE ions. The probe laser beam was focused on the sample with a lens of 75-mm focal length. The laser light power, in these measurements, was usually $\sim$ 12 mW that corresponded to the light power density around 12$\cdot 10^2$ W/cm$^2$. Dependence of the polarization noise power, normalized to that of the shot noise, at these intensities, was found to be linear, which indicated negligibly small role of optical nonlinearity. Polarization noise of the laser beam, transmitted through the sample, was analyzed using a spectrum analyser RSA6100A (Tektronix) in the frequency range up to 1~GHz (limited by the bandwidth of the balanced detector).

We did not expect to find any pronounced magnetic-field-dependent structure of the polarization noise spectrum, and, for this reason, we have developed a special procedure of the signal normalization that did not imply magnetic-field switching often employed in the spin-noise experiments\cite{muller}. To obtain calibrated value of the polarization noise power, we performed two measurements of the signal with the polarizer (LP) placed before and after the sample (positions (1) and (2) in Fig.~\ref{setup}), and one more measurement with blocked light. The results of the first two measurements ($S$ and $N$) differed only by the value of the polarization noise, and the third measurement provided the value of the dark noise ($E$). The final spectrum of the noise signal was calculated using the formula $\frac{S-N}{N-E}$, which allowed us to get rid of the influence of the frequency response function of the detector and to obtain directly the spectrum of the polarization noise power normalized to the shot-noise level.

\section{Specific features of the noise.}
The Nd- and Yb-doped glasses were studied, respectively, in the range of transitions 4I$_{9/2}$\,--\,4F${_3/2}$ ($\lambda \approx$ 870 nm) and $^2$F$_{7/2}$\,--\,$^2$F$_{5/2}$ ($\lambda \approx$ 980 nm), accessible for the Ti:sapphire laser. In both samples, we have found a fairly strong polarization noise, comparable in magnitude with the shot noise power, which was definitely related to the impurity ions, but could not be consistently explained in terms of spin fluctuations observed in RE-doped crystals\cite{kamenskii}.

\begin{figure}
\centering
 \includegraphics[width=0.7\linewidth]{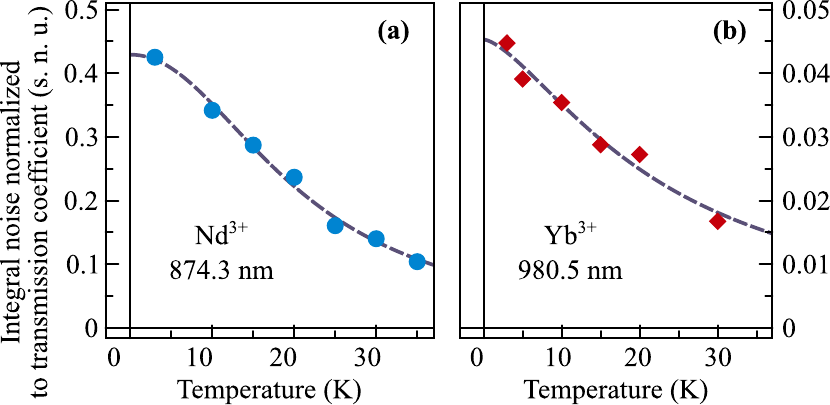}
 \caption{Temperature dependences of the polarization-noise power for (a) the Nd- and (b) Yb-doped glasses. Dashed lines are the power-law fits by the formulas $(a+bT^n)^{-1}$ with $n =$ 2.1 and 1.5 for the two glasses, respectively.}
 \label{temp} 
\end{figure}

Optical spectra of the polarization noise were measured, point-by-point, by integrating the noise signal over the frequency range 2 -- 220 MHz. The spectral dependences thus obtained, overall, well correlated with the polarization-noise gain effect that predicted high sensitivity of the probe beam polarization to fluctuations of individual homogeneously broadened resonances (provided that they are sufficiently narrow). Indeed, polarization noise was observed only in the region of absorption of the ions and, for both samples, was shifted towards longer wavelengths (Fig.~\ref{temp}). This fact also looks natural: a faster (nonradiative) ladder-type relaxation from the upper Stark sublevels broadens their homogeneous width and thus reduces the signal. In the Nd-doped glass, where the two crystal-field component of the excited multiplet are well resolved, this effect is especially pronounced -- the noise is observed only on one of the components. In the Yb-doped glass, no splitting is seen in the optical spectrum, and this effect is revealed just as a long-wavelength shift of the polarization noise spectrum with respect to the inhomogeneously broadened absorption line. 

\begin{figure}
\centering
\includegraphics[width=0.7\linewidth]{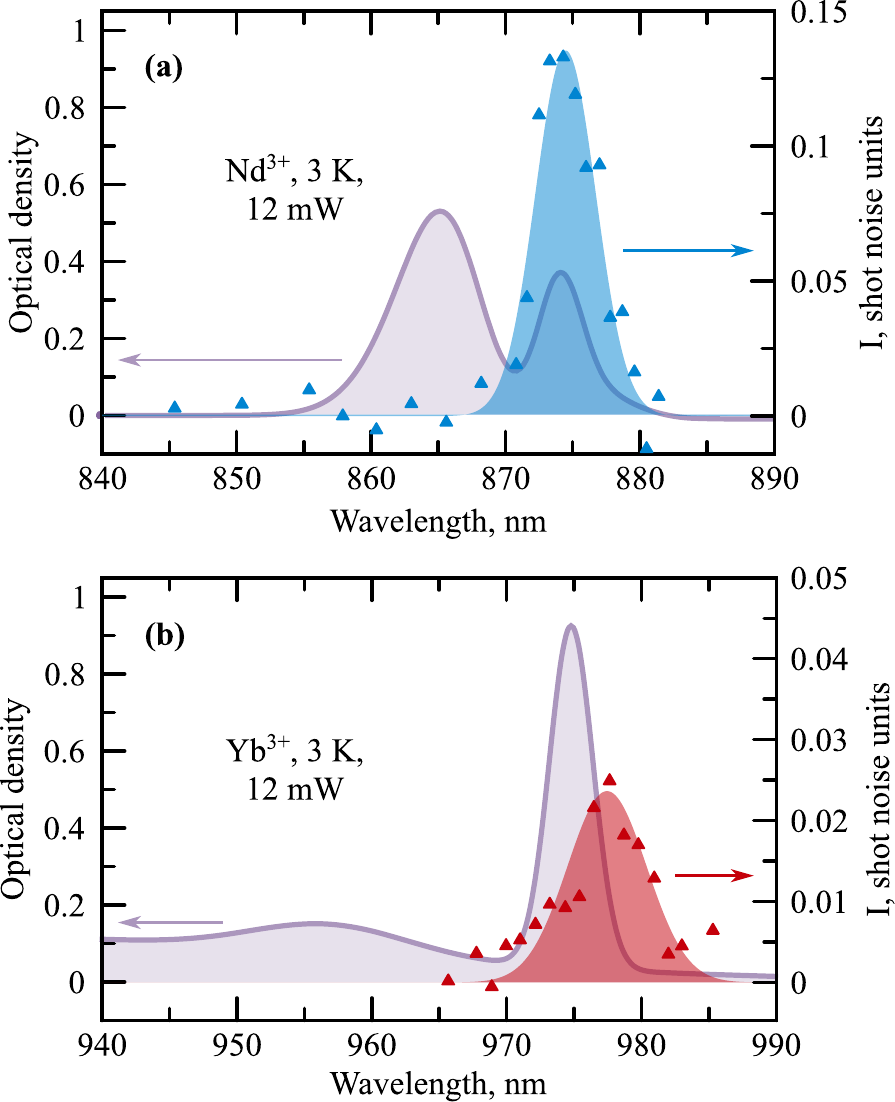}
\caption
{Optical spectra of absorption and polarization noise power (painted blue and red) for the samples of Nd- and Yb-doped glasses ((a) and (b), respectively). }
 \label{wl} 
\end{figure}

\begin{figure}
\centering
 \includegraphics[width=0.7\linewidth]{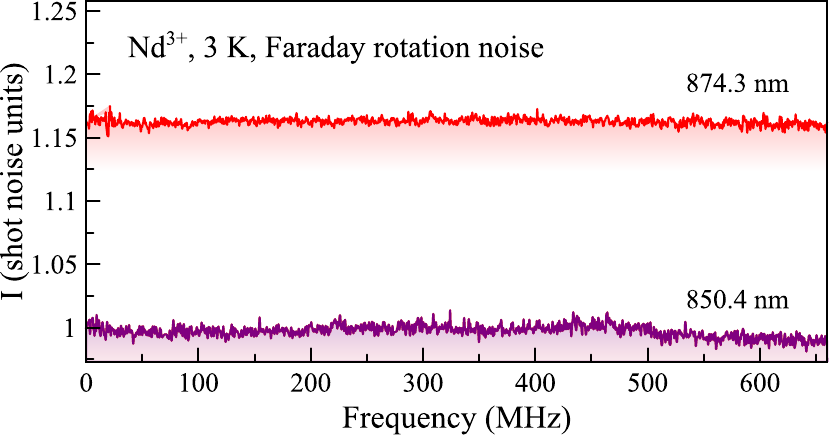}
 \caption{The polarization-noise spectra of the Nd-doped glass at two wavelengths of the probe light corresponding to the regions of absorption and transparency of the sample. At the wavelength of transparency ($\lambda =$ 850.4 nm), where no polarization noise is observed (see Fig.~\ref{temp}), the noise power normalized to the shot-noise level equals unity.The figure also demonstrates flatness of the polarization-noise spectrum. 
} 
\label{nd-white}
 \end{figure}

Temperature dependence of the noise (Fig.~\ref{nd-white}) also qualitatively agrees with the model of the polarization noise gain in an inhomogeneously broadened optical transition: with increasing temperature, the homogeneous width of the optical transition increases, and the polarization-noise power, controlled by the noise gain effect, decreases. Moreover, this temperature dependence allows one to restore, more accurately, temperature behavior of the homogeneous linewidth. Numerous experimental and theoretical studies indicated the power-law temperature dependence (often close to quadratic) of the homogeneous linewidth $\delta$ of optical transitions in RE-doped glasses \cite{shelby,brundage,macfarlane,vanderzaag,mins,bigot}: $\delta=\alpha+\beta T^n, n\sim 2$. Correspondingly, temperature dependence of the polarization noise power $\cal N$, scaling as inverse homogeneous linewidth of the optical transition (${\cal N}\sim 1/\delta$), should have the form:
\begin{equation}
 {\cal N}=[ a+bT^n]^{-1}
\label{1}
\end{equation}

The experimental temperature dependences were fitted by this formula and the obtained values of $n$ (2.1 for the Nd- and 1.5 for Yb-doped glasses) were found to be close to those obtained on similar samples using the technique of fluorescence line narrowing (2.4 and 1.3, respectively)\cite{brundage,brundage2}. 

Other properties of the observed polarization noise, however, convincingly showed that it cannot be associated with spin noise of the rare-earth ions. Specifically, the polarization-noise spectrum, within the range of 0.1\ldots1 GHz was practically ‘white’ (Fig.~\ref{nd-white}) This picture gives the idea of the magnitude of the polarization noise and the signal-to-noise ratio in these measurements. It is important that the polarization noise signal did not vary with magnetic field at least up to 0.4 T. One can easily estimate that in the field of 0.4~T, all reasonable precession frequencies of Nd$^{3+}$ ions should be higher than 100 MHz. It is also noteworthy that magnitude of the discovered polarization-noise power is huge as compared with that of spin noise in the RE-doped crystals, where the noise power of smaller magnitude (a few percent of shot-noise level) is concentrated in a much smaller frequency range ($\sim$ 30 MHz).
 
\begin{figure}
\centering
 \includegraphics[width=0.7\linewidth]{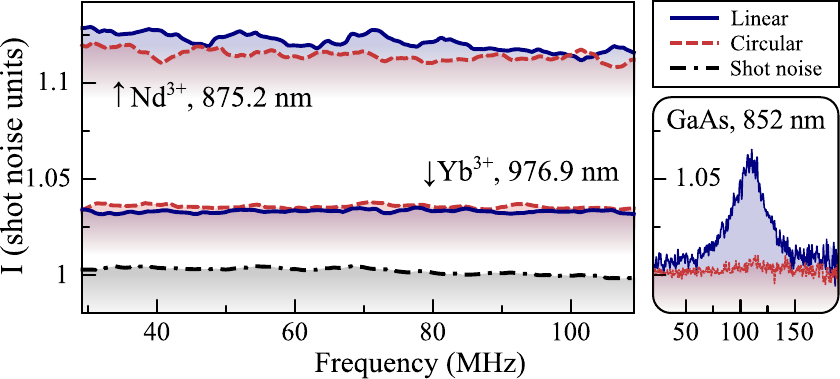}
 \caption{Comparison of the polarization-noise spectra for the Nd- and Yb-doped glasses obtained in the linearly and circularly polarized light. For the fluctuating gyrotropy, the noise in a circularly polarized light should vanish as shown in the right panel for the Faraday-rotation noise in n-GaAs (the latter experiment was performed specially to confirm this fact).}
 \label{circ-lin} 
 \end{figure}

We have also performed measurements of polarization noise, on the same experimental setup, in a circularly probe laser beam, when, as was noted in Sec.1, the noise of gyrotropy should not be observed. In this scheme, the DC signal of the balanced detector is always zero (the detector does not need to be balanced), and the measurements may be performed at any orientation of the half-wave plate 
 before the detector. These measurements have shown that the noise signal, within the experimental error, remained the same (equal to that in the linearly polarized probe, see Fig.~\ref{circ-lin}). Thus, we may conclude that the observed polarization noise is related to fluctuations of {\it linear {anisotropy}} rather than 
 fluctuations of gyrotropy. In combination with the above-mentioned high integrated power of the noise and its insensitivity to the external magnetic field, we can conclude that the observed polarizarion noise is not related to spin fluctuations of the impurity ions in glasses. 
 {}

\section{The nature of the noise}

Energy-level structure of trivalent RE ions in the visible spectral range is governed by spin-orbit splitting of intraconfigurational 4\textit{f}$^n$ - 4\textit{f}$^n$ transitions with well-defined energies. In crystalline matrices, energy levels of the ionic states exhibit additional splitting, which is the same for identical sites of the ion in the lattice. As a result, optical spectra of \textit{f-f} transitions in crystals are fairly narrow (fractions of wavenumbers). In glasses, local environments of different ions is different, and absorption spectra of \textit{f-f} transitions appear to be strongly broadened {\it inhomogeneously}. It is important that local environments of the ion differ from site to site not only by strength of the local field, but also by its symmetry, and optical transitions between the same multiplets of the ion differ not only in frequency, but also in optical anisotropy. 
Unlike crystals, glass are characterized by a great amount of equilibrium atomic configurations with nearly equal energies. As a result, glasses reveal, in the low-frequency region (below 10$^{10}$ Hz), a continuous spectrum of states responsible for anomalous behavior of their heat capacity and thermal conductivity \cite{phil1}. 
 Theoretical description of this atomic motion usually performed in terms of tunneling two-level systems (TLS), show that their density of states does not vanish at zero frequencies and is characterized, at low frequencies, by a `white' spectrum.

Now, turning to the discovered polarization noise we should note, first of all, that this kind of noise, as far as we know, has never been directly observed in regular crystalline structures and, is, therefore, specific for disordered media. Then, the fluctuating motion of the RE-ion environment should produce fluctuations of the polarizability tensor of the ion. Generally, such rearrangements of charges surrounding the impurity ion should modulate symmetric part of its polarizability tensor and should not affect its gyrotropy, which is observed experimentally. 

It is essential that homogeneous width of optical transitions in glasses with RE ions is relatively small ($\sim $ 10$^6$ - 10$^8$~Hz), so that fluctuations of the local field strength may shift the spectral line by more than (or comparable with) its half-width. It means that polarization of the near-resonant probe beam can be highly sensitive to these fluctuations. In fact, the mechanism of high sensitivity of the probe beam polarization to fluctuations of a quantum system is identical to that of the ‘giant spin noise gain effect’, described in \cite{kamenskii}, with the only difference that the spin dynamics of a paramagnetic ion is replaced by dynamics of its local environment. Efficient conversion of structural fluctuation of the glass into polarization noise of the probe beam can be observed only under resonant probing which is confirmed in our experiments. Spectral dependences of the polarization noise power presented above support the role of the homogeneous width of optical transition in the noise formation. 

 Thus, in our opinion, all the experimental data unambiguously support the proposed direct relationship between the observed polarization noise of the RE-doped glasses and structural dynamics of the vitreous medium. 

\section{Concluding remarks}
In this work, we applied the polarization-noise spectroscopy to glasses doped with RE ions and have found a strong stochastic signal resulted from fluctuations of linear {anisotropy} of the impurity centers induced, in turn, by fluctuating rearrangement of the RE-ion local environment. The noise power, in the applied method, is directly related to homogeneous width of the probed optical transition and can be used for its evaluation. 

It should be noted that temperature behavior of homogeneous linewidth of optical transitions of RE ions in glasses has been studied in numerous experimental works and was unambiguously ascribed to perturbations of the impurity ions by two-level systems of the vitreous matrix~\cite{vanderzaag,lyo,huber,silbey,schmidt,bai,feofil}. All the measurements of the homogeneous linewidth required `penetration' into the inhomogeneously broadened spectral line and used the effects of nonlinear optics such as spectral hole burning~\cite{kurita, bigot}, fluorescence line narrowing~\cite{brundage,shelby,bigot4}, or photon-echo effects~\cite{hegarty,ding}. 

The approach to these measurements proposed in the present work is essentially different. Its most important distinction is that it does not imply any nonlinearity of the medium and that the spectral packet of homogeneous width is selected, in the optical spectrum, {\it spontaneously} due to uncorrelated fluctuations of neighboring packets. It is also important that the noise signal arising under the above conditions in glasses, like it occurs in spin noise spectroscopy, is not a {\it response} to an external perturbation. As a result, the rules of the theory of linear response are violated, and the optical spectroscopy of polarization noise acquires features more typical for nonlinear optics\cite{glazov}. We believe that novelty of this measuring procedure and novelty of the discovered effect will open new ways of research in the high-resolution optical spectroscopy. The most promising application of this effect, in our opinion, should be based on quantitative analysis of the polarization noise power in different glass matrices with different RE impurities. It is noteworthy that these experimental approach can be also applied to glasses with other types of impurities with sufficiently narrow homogeneous widths of optical transitions (like transition metal ions). Altogether, we have shown that impurity ions with small homogeneous width can be used as sensors of structural low-temperature dynamics of disordered matrices. 
 
\section*{Acknowledgements}

The work was fulfilled under financial support of the Russian Science Foundation Grant No.~21-72-10021. The samples preparation was supported by Saint Petersburg State University (Grant No. 94030557).





\end{document}